\documentclass{ws-procs961x669}            
\begin{document}
\title{
Gravitational fields with sources: From compact objects to black holes\footnote{MG14 - BH4 Parallel Session}}

\author{Jos\'e~P.~S.~Lemos$^1$, Paolo Pani$^{2,1}$}

\address{$^1$Centro Multidisciplinar de Astrof\'{\i}sica - CENTRA, Departamento de F\'{\i}sica, Instituto Superior T\'ecnico, Universidade de Lisboa, Avenida~Rovisco Pais 1, 1049 Lisboa, Portugal.\\
$^2$Dipartimento di Fisica, ``Sapienza'' Universit\`a di Roma \& Sezione INFN Roma1, Piazzale Aldo Moro 5, 00185, Roma, Italy.\\
}

%

\begin{abstract}
We report on the Parallel Session BH4 \emph{``Gravitational fields with sources: From compact objects to black holes''} of the 14th Marcel Grossmann Meeting held at Sapienza University of Rome in 2015.
\end{abstract}


\bodymatter

\section{Introduction}\label{sec:intro}

In its 100-year-long history, general relativity (GR) has passed many stringent tests and is now accepted as the standard theory of gravity and one of mankind's greatest achievements. Most experiments performed in the last century had probed only the weak-field, quasi-Newtonian regime of the theory. This state of affairs has changed completely in the last months, after aLIGO's gravitational-wave (GW) detection of a black-hole (BH) binary coalescence~\cite{Abbott:2016blz}, which has finally given us access to the strong-field, highly-dynamical regime of Einstein's theory. In this regime the structure and dynamics of compact objects and BHs might differ from GR with dramatic and potentially observable effects. 

Incidentally, the BH4 Session\footnote{\url{http://centra.tecnico.ulisboa.pt/network/grit/mg14/}} of the 14th Marcel Grossmann Meeting was devoted to recent theoretical developments on one of the most important aspects of the newly-born GW era, namely, on the physics of compact objects in Einstein's gravity and in modified theories of gravity. The topics covered in the session were diverse and include compact objects as probes of fundamental physics, regular BHs, quasiblack holes, wormholes, tests of gravity with BHs and neutron stars (NSs), exact solutions, BHs with hair, and strong-gravity effects in NSs beyond GR.

Given the interdisciplinary nature of the subject some overlap exists with other sessions of the
meeting, in particular with the AT sessions (Alternative Theories), some other BH sessions (Black Holes: Theory and Observations/Experiments), the BS sessions (Boson Stars), some of the GW sessions (Gravitational Waves) and some of the NS sessions (Neutron Stars). We invite the interested reader to see also the reports of these other sessions to get a complete overview of the state of the art in these areas.

In view of the multitude of covered topics, we find it useful to divide this report in separate sections, which reflect the
internal schedule that was adopted during the meeting (we refer to the webpage \url{http://centra.tecnico.ulisboa.pt/network/grit/mg14/} for details).

\section{Tests of GR and fundamental fields with BHs}

A central topic of the session was related to BH-based tests of gravity and fundamental fields (cf. Refs.~\cite{Yunes:2013dva,Brito:2015oca,Berti:2015itd,CQGFocus,Yagi:2016jml} for some reviews on this topic).

Cosimo Bambi gave an overview on electromagnetic tests of the Kerr paradigm, i.e. the GR prediction that astrophysical BHs are described by the Kerr metric. The properties of the electromagnetic radiation emitted by the gas in the accretion disk can be potentially used to test the geometry of the spacetime around dark compact bodies~\cite{Bambi:2015kza}. Bambi focused on the study of the thermal spectrum of thin disks and on the analysis of the iron K-alpha line. Both techniques can already be used to provide some constraints on possible deviations from the Kerr solution, but more interesting tests require new X-ray facilities with a larger effective area.

Various talks of this part of the session were devoted to BH superradiance~\cite{Teukolsky:1974yv} and to the BH superradiant instability (for a recent review see~\cite{Brito:2015oca}). Richard Brito gave an overview on this topic, especially on the superradiant instability of spinning BHs in the presence of ultralight bosons\cite{Arvanitaki:2010sy,Pani:2012vp,Pani:2012bp,Witek:2012tr,Brito:2013wya}. This process has been used to impose strong constraints on ultralight dark-matter candidates, thus turning BHs into effective particle detectors~\cite{Brito:2015oca,CQGFocus}. Brito described recent developments~\cite{Brito:2014wla} in this topic, including the impact of GW emission and gas accretion on the evolution and final state of the instability.

Jo\~ao Rosa presented his recent results~\cite{Rosa:2015hoa} for the superradiant wave scattering in a binary system that includes a spinning BH. He considered the case of magnetic dipole and gravitational quadrupole radiation emitted by a rotating NS orbiting a Kerr BH and showed that both electromagnetic and gravitational radiation can undergo superradiant scattering. For a co-rotating binary, amplification may occur when the pulsar's orbit is sufficiently inclined with respect to the BH equatorial plane. This effect leads to a peculiar modulation of the pulsar's luminosity that may potentially be used to test the occurrence of superradiance scattering, yielding the first example of an astrophysical system where this important effect may be probed observationally.

Juan Carlos Degollado discussed the GW signatures of massive scalar fields around BHs~\cite{Barranco:2012qs,Degollado:2014vsa}. The GW signal of a perturbed BH is affected by quasibound configurations of scalar fields, which are also responsible for the BH superradiant instability discussed above. By solving the Teukolsky equation for gravitational perturbations of a BH coupled to the Klein-Gordon equation, he discussed a characteristic GW signal, composed by a quasinormal ringing followed by a late time tail. This tail contains small-amplitude wiggles at the frequency of the dominating quasibound scalar state.

Related to this topic, Mohamed Ould El Hadj discussed the excitation of quasibound states and quasinormal modes of the Schwarzschild BH by a plunging particle coupled to a massive scalar field~\cite{Decanini:2014bwa,Decanini:2015yba}. He computed the waveform produced by a particle plunging from slightly below the innermost stable circular orbit into a Schwarzschild BH and analyzed its spectral content. By using a toy model in which the particle is linearly coupled to a massive scalar field, he discussed some important features of the process which could be helpful to interpret the results of dynamical processes involving BHs in other theories with massive bosons, most notably in massive gravity.

David Dempsey presented a recent paper with Sam Dolan, in which they discuss in details the bound states of the Dirac equation on Kerr spacetime~\cite{Dolan:2015eua}. Similarly to the case of massive bosons, also Dirac particles support (quasi)bound states, namely gravitationally-trapped modes which are regular across the (future) event horizon. However, at variance with the case of bosonic fields, Dirac bound states decay with time due to the absence of superradiance in the (single-particle) Dirac field~\cite{Brito:2015oca}. Dempsey presented a practical method for computing the spectrum of energy levels and decay rates, which agrees very well with known asymptotic results.

A further part of this session was devoted to tests of modified gravity~\cite{Yunes:2013dva,Berti:2015itd}.
Hajime Sotani discussed the prospect of constraining a particular extension to GR, namely Eddington-inspired Born-Infeld gravity~\cite{Banados:2010ix}, using terrestrial measurements of the neutron skin thickness of ${}^{208}$Pb together with the astronomical observations of the radii of light NSs~\cite{Sotani:2014goa,Sotani:2014xoa}.

Bethan Cropp discussed the concept of universal horizons in Lorentz-violating theories of gravity~\cite{Blas:2011ni,Barausse:2011pu}. In these theories superluminal motion alters the very notion of a BH. Nevertheless, in both Einstein-Aether gravity, and Horava-Lifshitz gravity, there is a causally disconnected region in the BH spacetime which traps excitations of arbitrarily high velocities. Cropp discussed some aspects of these ``universal horizons'', especially the role of their surface gravity and the study of the trajectories in these spacetimes~\cite{Cropp:2013sea}.

Finally, Paolo Pani presented recent progress in the theory of the tidal deformability of spinning compact objects. He presented powerful perturbative techniques to compute the geometry of a tidally-distorted, slowly-spinning compact object. The spin of the object introduces couplings between electric and magnetic deformations~\cite{Pani:2013pma} and new classes of induced Love numbers emerge. By applying this framework, he showed that the tidal Love numbers of a Kerr BH are zero generically to first order in the spin~\cite{Landry:2015zfa} and also to second order~\cite{Pani:2015hfa}, at least in the axisymmetric case, and that spin-tidal couplings can introduce important corrections to the gravitational waveforms of spinning NS binaries approaching the merger~\cite{Pani:2015nua}.

\section{Collisions and collapse}


The second part of the BH4 Session was devoted to dynamical processes related to BHs, in particular to the study of the gravitational collapse, scattering processes, BH perturbations and GW emission.

Reinhard Meinel presented a study to support the cosmic censorship conjecture in GR. He considered the quasi-stationary collapse of a uniformly rotating disc of dust~\cite{Kleinwachter:2010di,Breithaupt:2015xva} and showed how the latter leads to the formation of a BH, rather that of a naked singularity, and discussed possible extensions of this framework.


Mandar Patil presented a study which is also related to the cosmic censorship conjecture. Namely, he compared the ultra-high energy collision of point particles near quasi-extremal Kerr BHs and the same process near ``superspinars'' (the latter are spinning objects described by a Kerr solution exceeding the angular-momentum bound and therefore possessing a naked singularity)~\cite{Patil:2010nt,Patil:2011yb}. Fine-tuning of the geodesic parameters of the colliding particles is necessary to achieve large center-of-mass energy in the case of BHs, while no such fine-tuning is necessary in the case of over-spinning Kerr geometries. Furthermore, the collision near superspinars requires much shorter time and it is much more energetic than the corresponding process near BHs. These results might be interesting to find smoking guns of putative overspinning compact objects~\cite{Patil:2015fua}.


In the context of dynamical processes which are also sources of GWs, Giuseppe d'Ambrosi described a special class of ballistic geodesics in Schwarzschild spacetime for extreme-mass ratio binaries~\cite{d'Ambrosi:2014iga}. These orbits are in one-to-one correspondence to stable circular orbits. By deriving the source terms of the Regge-Wheeler and Zerilli-Moncrief equations for these orbits, he computed the GWs emitted during the infall of a point particle. Finally, he presented the resulting waveforms and discussed some limits and applicability of this approach.


Hirotada Okawa presented some recent numerical work on the gravitational collapse in confined geometries~\cite{Okawa:2014nea,Okawa:2015xma,Okawa:2013jba}, which is motivated by the nonlinear instability of anti-de Sitter spacetime against gravitational collapse~\cite{Bizon:2011gg}. Small perturbations in confined geometries such as anti de Sitter do not simply decay away since they can be reflected by the boundary of the spacetime. Such waves can interact nonlinearly producing turbulence, i.e. the energy is transferred to ever higher frequency modes and is confined to ever smaller spacetime regions, eventually resulting in the formation of a BH. Okawa presented the results of some numerical simulation of confined Einstein-scalar theory~\cite{Okawa:2014nea,Okawa:2015xma,Okawa:2013jba}, including the appearance of long-lived bound states, ``islands of stability'' in the parameter space of the initial data, and he highlighted the generic role of confinement for the nonlinear instability.


Sabbir Rahman presented a work in which he argued that the formation of a Schwarzschild BH via Oppenheimer-Snyder type gravitational collapse must be accompanied by a change in topology upon formation of the event horizon which physically separates matter in the interior from that of the exterior~\cite{Rahman:2015lba}. He described that ---~while collapsing matter crossing the event horizon continues to fall towards the singularity of the Schwarzschild interior~-- this region does not in fact contain the matter originally responsible for the collapse. Rather, the latter occupies a distinct internal spacetime region with its own independent evolution. Some interesting implications of this analysis for what concerns the interior region of a Schwarzschild BH were also discussed.


For what concerns scattering processes in BH spacetimes, Luis Carlos Crispino presented some results on the scattering cross section of Reissner-Nordstr\"om BHs for the case of a pure electromagnetic incident wave~\cite{Crispino:2009ki,Crispino:2015gua}. He described how scattering is affected both by the conversion of electromagnetic to gravitational radiation, and by the parity-dependence of the phase shifts induced by the BH charge. The latter effect creates a helicity-reversed scattering amplitude that is non-zero in the backward direction. Interestingly, from the features of the electromagnetic wave scattered in the backward direction it is possible to tell if a static BH is charged or not~\cite{Crispino:2014eea}.


Pavel Spirin discussed an analysis~\cite{Spirin:2015wwa} on the angular and frequency characteristics of the gravitational radiation emitted in collisions of massless particles in the context of classical GR for small values of the Schwarzschild radius to impact parameter ratio, $\alpha$. He presented a comparison with previous results in the literature and showed that the radiation efficiency outside a narrow cone of angle $\alpha$ in the forward and backward directions with respect to the initial particle trajectories is $\sim \alpha^2$.


Concerning the scattering coefficients of linear perturbations around BHs, F\'abio Novaes computed their exact form using monodromy data of the associated radial wave equation. He applied a recent isomonodromic approach~\cite{Novaes:2014lha} to Kerr-NUT (anti) de Sitter BHs, whose wave equation is separable. The nontrivial monodromy is obtained by numerical integration, whereas in the near-extremal case an approximate analytical expression was also found. Finally, these results were discussed in the context of BH superradiance and BH quasinormal modes in four and higher dimensions.


Mengjie Wang presented a detailed analysis of Maxwell perturbations in Kerr-anti de Sitter BHs~\cite{Wang:2015fgp,Wang:2015goa}. The boundary conditions for these perturbations are often obtained by imposing a vanishing field (Dirichlet condition) at the anti de Sitter boundary. While such conditions imply a vanishing energy flux at the boundary, the converse is generically not true. Wang argued that a vanishing energy flux is a more fundamental physical requirement and showed that the corresponding boundary conditions (of Robin type) can lead to a new branch of quasinormal modes of Maxwell perturbations on a Kerr-anti de Sitter BH. Finally, he reported on some results on the superradiant instabilities and on vector clouds in Kerr-anti de Sitter spacetime.


Concerning the BH superradiant instability and bosonic clouds around BHs~\cite{Brito:2015oca}, Hirotaka Yoshino gave an overview on his  work~\cite{Kodama:2011zc,Yoshino:2012kn,Yoshino:2013ofa,Yoshino:2014wwa,Yoshino:2015nsa} on the dynamics of a string axion field around a spinning BH. He first revisited the calculation of the growth rate of the superradiant instability and showed that, in some cases, overtone modes have larger growth rates than the fundamental mode with the same angular quantum numbers when the BH is rapidly rotating. Furthermore, he presented the dynamical evolution of the scalar field with nonlinear self-interactions, and discussed the dependence of the dynamics on the axion mass. These results suggest that a fairly strong GW burst is emitted during the bosenova~\cite{Yoshino:2012kn,Yoshino:2015nsa}. This signal might be detectable by the ground-based interferometers if it happens in our Galaxy or in nearby galaxies.


Finally, Daniele Guariento presented an analysis of the causal structures of cosmological BHs undergoing scalar-field accretion~\cite{Afshordi:2014qaa,daSilva:2015mja}. He showed that the generalized McVittie solution ---~which describes a central time-dependent mass in an expanding cosmological background~--- is an exact solution of a self-gravitating subset of Horndeski scalar-tensor theories (see~\cite{Berti:2015itd} for a review), and constitutes an example of an analytic solution of hairy, time-dependent BHs. He demonstrated that a time-dependent central mass may have a significant impact on the overall causal structure of the spacetime. The metric always has an event horizon at future cosmological time infinity in the appropriate limits, but the character of the horizon depends on the accretion and cosmological histories in the bulk~\cite{Afshordi:2014qaa}. 
%

\section{Compact stars and BH solutions, entropy and thermodynamics}

The first part of the second day of the BH4 Session started with a number of contributions on novel self-gravitating compact solutions in Einstein theory coupled to different matter fields.

%

Carlos Herdeiro presented an overview of different BH solutions with scalar hair in asymptotically flat spacetime. These solutions were presented in different classes according to the assumptions of the no-scalar-hair theorem that they violate. Particular emphasis was given to the solution found
by Herdeiro and Radu describing a Kerr BH with scalar hair, which solves Einstein gravity minimally coupled to a massive complex scalar field~\cite{Herdeiro:2014goa,Herdeiro:2015gia} (see~\cite{Herdeiro:2015waa} for a review). In this solution the complex scalar field oscillates in time but the metric remains stationary and axisymmetric. 
Herdeiro discussed the connection between these solutions and the superradiant instability, as well as some phenomenological properties of the geometry.


Related to the previous topic, Marco Sampaio presented a study~\cite{Sampaio:2014swa} on marginally-stable bosonic clouds in Reissner-Nordstr\"om spacetimes made of massive scalar and massive vector (Proca) fields. These configurations exist at the threshold of the superradiant instability and often have a nonlinear counterpart similarly to the above nonlinear solution discussed by Herdeiro for the case of Kerr BHs. In the work presented by Sampaio, the authors computed the quasi-bound state frequencies and established that no bound states exist in this spacetime. Nevertheless, they found nontrivial configurations with an arbitrarily small imaginary part of the frequency (outside the superradiant regime), which can be arbitrarily long lived. An interesting open question is related to the possibility of finding nonlinear counterparts to such linear configurations.


Vilson Zanchin discussed some regular, exact BH solutions of GR coupled to Maxwell's electromagnetism and charged matter~\cite{Guilfoyle:1999yb}. These solutions represent spherically symmetric charged perfect fluid distributions whose metric potentials and electromagnetic fields are related in some particularly simple form. He showed that, for certain range of the parameters of the model, there are objects which correspond to regular charged BHs, whose interior region is filled by a charged phantom-like fluid, or, in the limiting case, de Sitter, and whose exterior region is Reissner-Nordstr\"om. There are several types of solutions: regular non-extremal BHs with a timelike smooth boundary and regular BHs with a null matter boundary. The main physical and geometrical properties of such charged regular solutions were analyzed~\cite{Lemos:2011dq,Lemos:2016ulj}.


Jos\'e Sande Lemos's talk was about the compactness of relativistic charged spheres. He presented a generalization of the Buchdahl bound on the compactness of a perfect-fluid star to the case of electrically charged stars due to Andr\'easson. This Buchdahl-Andr\'easson bound is found through the assumption that the radial pressure plus twice the tangential pressure of the matter is less than the energy density. A class of configurations that saturate the electrically charged Buchdahl-Andr\'easson bound are electrically charged shells. Another class of configurations was presented by Lemos in his contribution~\cite{Lemos:2015wfa}. Indeed, Guilfoyle's electrically charged stars which have a very stiff equation of state ---~namely the Cooperstock-de la Cruz-Florides equation of state~--- also saturate the bound. When the electric charge is zero Guilfoyle's stars reduce to the Schwarzschild incompressible stars. An interesting open problem is to find a proof ``a la Buchdahl'' such that these configurations are also the limiting configurations of the Buchdahl-Andr\'easson bound.


Jos\'e Arba\~nil presented a work~\cite{2015PhRvD..92h4009A} in which he studied the hydrostatic equilibrium and the stability against radial perturbations of charged strange quark stars composed by a charged perfect fluid. In his model the perfect fluid follows the MIT bag model equation of state and the charge has a power-law distribution in the radial coordinate. He found that the total charge that affects appreciably the stellar structure is very large (about $10^{20}$ Coulomb) and that, for some range of the parameters, the electric charge contributes to the stability of the star. Finally, he discussed the necessary and sufficient conditions for the stability against radial oscillations in terms of the maxima of the stellar mass as a function of the central density for fixed electric charge.


T\'erence Delsate presented a work done in collaboration with Robert Mann in which they obtained an exact BH solution whose thermodynamical properties are those of a van der Waals fluid~\cite{Delsate:2014zma,Rajagopal:2014ewa}. These BHs are discussed for arbitrary number of dimensions and various horizon geometries. These solutions are asymptotically anti de Sitter and the cosmological constant is treated as a pressure in the so-called extended phase space. Delsate discussed the properties of the effective source of these BHs, as well as the energy conditions of the matter fields.


Finally, this part of the session terminated with three talks on the thin shells in BH spacetimes. Paulo Luz discussed the junction of an interior Minkowski metric with an exterior Reissner-Nordstr\"om spacetime by using the Israel-Darmois formalism. After defining the Kruskal-Szekeres coordinates for the maximally extended Reissner-Nordstr\"om spacetime, he discussed the properties of the thin shell, such as its energy density and pressure, as a function of its position. Finally, Luz also analyzed the energy conditions verified by a thin shell for all possible cases~\cite{Lemos:2016xxx}.


Masato Minamitsuji investigated the thermodynamic equilibrium states of a rotating thin ring shell in a ($2+1$)-dimensional spacetime with a negative cosmological constant~\cite{Lemos:2015zma}. Inside the ring, the spacetime is pure anti-de Sitter, whereas the exterior metric is described by the Ba\~nados, Teitelboim and Zanelli (BTZ) spacetime. The first law of thermodynamics of the thin shell, with three equations of state for the pressure, angular velocity and local temperature of the shell, leads to the entropy of the shell which only depends on its gravitational radius. In the limit in which the shell location coincides with its own gravitational radius, the entropy of the shell coincides with the Bekenstein-Hawking entropy.


On a related topic, Gon\c{c}alo Quinta studied a $d$-dimensional static thin shell in a Schwarzschild spacetime from a mechanical and a thermodynamical point of view~\cite{Lemos:2015gna}. By specifying the nature of the spacetime around the shell he found a set of equations for the rest mass density and pressure which determine the mechanical behavior of the shell. A thermodynamic study of the solution can be performed by first calculating the entropy of the shell and then by analyzing its thermodynamic stability. Again, when the shell is located at its gravitational radius, one recovers the Bekenstein-Hawking entropy of a $d$-dimensional Schwarzschild BH.

\section{Tests of GR and fundamental fields with compact objects}

Following the first part of the BH4 Session, the last part was devoted to tests of gravity and fundamental fields using compact objects rather than BHs~\cite{Berti:2015itd}.

%

Carlos Palenzuela presented his recent work in collaboration with Steven Liebling on some new constraint on scalar-tensor theories from the most massive NSs~\cite{Palenzuela:2015ima}. Although severe constraints on these theories are in place (mostly from solar system experiments and binary-pulsar observations), there remains a large set of scalar-tensor theories which are still consistent with all observations. Palenzuela discussed how to constrain the viable region of the parameter space through a stability analysis of highly compact NSs in these theories. He presented a numerical evolution of very compact stars and identified under which conditions these stars are unstable against collapse to a BH. Finally, he discussed the implications of these results for the most recent bounds on the mass and radius of isolated NSs.


Kalin Staykov presented recent work on neutron and strange stars in a particular $f(R)$ theory of gravity, namely the so-called $R$-squared gravity, where $f(R)=R+aR^2$ and $a$ is an arbitrary parameter~\cite{Yazadjiev:2014cza,Staykov:2014mwa,Staykov:2015cfa}. He discussed slowly rotating models as well as f-mode oscillations~\cite{Kokkotas:1999bd} and compared these results to the case of GR. Finally, Staykov presented some asteroseismology relations which are almost insensitive to the equation of state and to the parameter $a$.


Sindy Mojica discussed a work done in collaboration with Burkhard Kleihaus and Jutta Kunz on the quadrupole moments of rapidly rotating compact objects in dilatonic Einstein-Gauss-Bonnet theory~\cite{Kleihaus:2014lba}. By solving numerically the field equations for a stationary axisymmetric object, she determined observables such as the mass, the angular momentum, the moment of inertia, and the quadrupole moment of BHs and NSs (for various equations of state) in this theory. In the NS case she found that the relation between the scaled moments of inertia and the scaled quadrupole moments is almost independent of the equation of state, when the scaled angular momentum is held fixed. These relations generalize the ``I-Love-Q'' relations found by Yagi and Yunes~\cite{Yagi:2013awa,Yagi:2013bca} to the case of dilatonic Einstein-Gauss-Bonnet theory.


Adolfo Cisterna presented his work done in collaboration with T\'erence Delsate and Massimiliano Rinaldi on stellar equilibrium configurations in a subset of Horndeski’s gravity characterized by a coupling between the kinetic scalar-field term and the Einstein tensor~\cite{Cisterna:2015yla}. He showed that, in a certain limit, there exist solutions that are identical to the Schwarzschild metric outside the star but change considerably in the interior, where the scalar field is nontrivial. He also discussed the region of parameter space where NSs can exist and compared the mass-radius relation of NSs in this theory with their GR counterpart.


Caio Macedo presented some recent work on slowly rotating anisotropic NSs in GR and in scalar-tensor theory~\cite{Silva:2014fca}. Some models of matter at ultranuclear density suggest that the core of a NS may be significantly anisotropic. Macedo discussed the effects of anisotropy on slowly rotating stars in GR and also in one of the most popular extensions to Einstein's theory, namely scalar-tensor theory. In the latter context, he studied the effect of the anisotropy on the spontaneous scalarization that might occur for compact stars in scalar-tensor theories. He found that the effects of scalarization increase (decrease) when the tangential pressure is bigger (smaller) than the radial pressure, and he presented a simple criterion to determine the onset of scalarization in this model. These results suggest that binary pulsar observations have the potential to constrain the degree of anisotropy in NS cores.


In the context of anisotropic stellar configurations, Mahsa Kohandel presented a recent work on the effects of an anisotropic pressure on the boundary conditions of the anisotropic Lane-Emden equation~\cite{Shojai:2015hsa}. She presented some new exact solutions of the anisotropic Lane-Emden equation and extended some theorems for Newtonian perfect fluid stars to the case of anisotropic fluids.


In another related contribution, Arman Stepanian presented the extension of the previous results to the case of anisotropic stars in GR, discussing static, spherically symmetric fluids with anisotropic pressure. He obtained some new classes of exact solutions of the Lane-Emden equations and discussed their properties.


Luis Gonz\'alez-Romero presented some universal relations for the quasinormal modes of realistic NSs containing exotic matter~\cite{BlazquezSalcedo:2012pd,Blazquez-Salcedo:2013jka}. He discussed both axial and polar modes and different equations of state, most of them containing exotic matter. He and collaborators obtained new phenomenological relations between the frequency and damping time of the modes, which are approximately independent of the equation of state and that relate the quasinormal mode spectrum of the star to its global properties. These relations could be used to constrain the equation of state of NSs with GW asteroseismology. Finally, he presented some quasi-universal relations between the quasinormal modes and the ``I-Love-Q'' parameters~\cite{Yagi:2013awa,Yagi:2013bca} for NSs with exotic matter.


Remo Garattini discussed compact stars in the context of so-called Gravity's Rainbow~\cite{Garattini:2015vam}. He showed that, under certain conditions, the energy density obtained by quantum fluctuations of the vacuum can be interpreted as a Dev-Gleiser model for a compact star. Finally, he also presented some application of this framework to the Tolman-Oppenheimer-Volkoff equations describing perfect-fluid relativistic stars.


Francisco Lobo presented a novel framework for describing a wide class of generic spherically symmetric thin-shell wormholes~\cite{Garcia:2011aa,Bouhmadi-Lopez:2014gza}. By using a cut-and-paste procedure, he analyzed the stability of arbitrary spherically symmetric thin-shell wormholes against radial perturbations. He demonstrated that the stability of the wormhole is equivalent to choosing suitable properties for the exotic material residing on the wormhole throat. As an application, he studied novel wormhole solutions supported by a matter content that minimally violates the null energy condition.


Finally, Oleg Zaslavskii presented recent developments of his recent work on ultra-high energy particle collisions near BHs~\cite{Zaslavskii:2011dz,Zaslavskii:2015ema}. He classified and discussed various types of collisions, including those occurring outside and inside the event horizon, the role of critical trajectories and of finite-size effects, the effects of the BH spin, electric charge and magnetic field, as well as some alternative processes related to spacetimes with ergoregions. He concluded that further studies of the Penrose process~\cite{Penrose:1969,Brito:2015oca} in combination with the Ba\~nados-Silk-West effect~\cite{Banados:2009pr} are needed to reach a better understanding of high-energy particle collisions near BHs.

\section{Conclusion}

The BH4 Session ``Gravitational fields with sources: From compact objects to black holes'' was characterized by very high and active participation and by a large number of speakers who delivered high-quality talks. Several interesting issues and ideas have been raised during the discussion.

The recent GW detection by aLIGO
promises to bring these areas of research to a new level, in which theoretical modeling can be finally compared to observations in the strong-field/highly-dynamical regime of gravity. As clear from the impressive list of contributions presented above, the community was already preparing for such revolution and we believe that several interesting results will appear in the near future. We hope to report on some of them at the 15th Marcel Grossmann Meeting in 2018.

\section{Acknowledgments}
This work was supported by FCT-Portugal through Projects 
No.~PEst-OE/FIS/UI0099/2014 and No.~IF/00293/2013 and by 
the COST Action MP1304 ``NewCompStar''.

\bibliographystyle{ws-procs961x669}
\bibliography{references}

\end{document}